\documentclass[a4paper,11pt]{article}
\usepackage{pos}
 \usepackage{delimset} 
 \usepackage{multirow}
 \usepackage{tikz}

\newcommand{\sfrac}[2]{{\textstyle\frac{#1}{#2}}}

\newcommand{\levz}{\mathrm{J}}

\newcommand{\superN}{\mathcal{N}}
\newcommand{\comm}[2]{[#1,#2]}
\newcommand{\tr}{\mathop{\mathrm{tr}}}
\newcommand{\levo}[1]{\mathrm{\widehat #1}}

\newcommand*\GG[5]{\,{}_{#1}F_{#2}\left[\genfrac{}{}{0pt}{1}{#3}{#4};#5\right]}
\newcommand{\dd}{\mathrm{d}}

\newcommand{\Eval}{s} 

\newcommand{\alg}[1]{\mathfrak{#1}}
\newcommand{\grp}[1]{\mathrm{#1}}
\newcommand{\gen}[1]{\mathrm{#1}}

\makeatletter
\def\mr@ignsp#1 {\ifx\:#1\@empty\else #1\expandafter\mr@ignsp\fi}%
\newcommand{\multiref}[1]{\begingroup
\xdef\mr@no@sparg{\expandafter\mr@ignsp#1 \: }%
\def\mr@comma{}%
\@for\mr@refs:=\mr@no@sparg\do{\mr@comma\def\mr@comma{,}\ref{\mr@refs}}%
\endgroup}
\makeatother

\newcommand{\hypref}[2]{\ifx\href\asklfhas #2\else\href{#1}{#2}\fi}
\newcommand{\Secref}[1]{Section~\multiref{#1}}

\renewcommand{\eqref}[1]{(\multiref{#1})}

\makeatletter
\newlength{\apb@width}
\newcommand{\autoparbox}[2][c]{\settowidth{\apb@width}{#2}\parbox[#1]{\apb@width}{#2}}
\newcommand{\includegraphicsbox}[2][]{\autoparbox{\includegraphics[#1]{#2}}}
\makeatother

\title{Massive Integrability: From Fishnet Theories to Feynman Graphs and Back}

\author*{Florian Loebbert}
\author[]{Julian Miczajka}

\affiliation[]{Institut f\"{u}r Physik, Humboldt-Universit\"{a}t zu Berlin, \\
Zum Gro{\ss}en Windkanal 6, 12489 Berlin, Germany}

\emailAdd{\{loebbert,miczajka\}@physik.hu-berlin.de}

\abstract{
An overview of the massive generalization of Yangian symmetry for Feynman integrals is given. We illustrate the relation to a massive fishnet theory defined as a double-scaling limit of Coulomb-branch $\superN=4$ SYM theory.
 }

\FullConference{%
  *** The European Physical Society Conference on High Energy Physics (EPS-HEP2021), ***\\
  *** 26-30 July 2021 ***\\
  *** Online conference, jointly organized by Universit\"at Hamburg and the research center DESY ***
}

\begin{document}
\maketitle

\section{Introduction}

In the context of the AdS/CFT correspondence, integrability has proven to be a useful tool.
Recently, it was found that large classes of Feynman integrals exhibit a conformal Yangian symmetry. Via their interpretation as correlation functions in the so-called fishnet theories, these integrable structures trace back to the integrability of massless planar $\superN$=4 super Yang-Mills (SYM) theory
\cite{Gurdogan:2015csr,Chicherin:2017cns,Chicherin:2017frs,Loebbert:2019vcj}.
In the following, we review the generalization to the massive case \cite{Loebbert:2020hxk,Loebbert:2020tje,Loebbert:2020glj}, which corresponds to the first findings of integrable structures in massive quantum field theory in four dimensions. 

\section{Yangian Symmetry and Integrability in AdS/CFT}

The Yangian algebra introduced by Drinfel'd  constitutes an infinite dimensional extension of a Lie algebra $\alg{g}$.
It underlies quantum integrable models with a rational R-matrix and plays an important role for the AdS${}_5$/CFT${}_4$ correspondence.
In its so-called first realization, the whole Yangian $Y[\alg{g}]$ is generated by two sets of generators. The level-zero generators of the Lie algebra $\alg{g}$ have a trivial tensor product action of the form $\gen{J}^a=\sum_{k=1}^n \gen{J}^a_k \in \alg{g}$ and obey the standard Lie algebra commutation relations
$ \comm{\gen{J}^a}{\gen{J}^b}=f^{ab}{}_c \gen{J}^c$.
In addition, the Yangian is generated by a set of level-one generators with a non-trivial coproduct:
\begin{equation}
\gen{\widehat J}^a=f^a{}_{bc} \sum_{j<k=1}^n \gen{J}_j^c \,\gen{J}_k^b+ \sum_{j=1}^n \Eval_j \gen{J}_{j}^a,  
\qquad\qquad
 \comm{\gen{J}^a}{\gen{\widehat J}^b}=f^{ab}{}_c \gen{\widehat J}^c.
 \label{eq:DefLev1}
\end{equation}
Moreover, the so-called Serre relations have to be obeyed, cf.\ Ref.\  \cite{JulianThesis} for the massive representation discussed below.
 In the case of $\mathcal{N}=4$ SYM theory, the underlying Lie algebra is $\alg{g}=\alg{psu}(2,2|4)$.
In the AdS/CFT context the Yangian can be understood as the closure of an  ordinary conformal symmetry and a dual conformal symmetry~\cite{Drummond:2009fd}.

\section{From $\mathcal{N}=4$ SYM to Massless Feynman Graphs}

A generalization of $\superN=4$ SYM theory is the so-called $\gamma$-deformation which is defined by introducing certain phase factors in the Lagrangian. These phases depend on three parameters $\gamma_j$ for $j=1,2,3$, which correspond to the three Cartan charges of the $\grp{SU}(4)$ R-symmetry. These parameters together with the coupling constant $g$ allow to take very interesting double-scaling limits of the $\gamma$-deformed theory \cite{Gurdogan:2015csr}. In the most restrictive limit with $g\to0$ and $\gamma_3\to i\infty$ one obtains a simple Lagrangian for two complex scalars $X$ and $Z$ (here we use Euclidean signature):
  \begin{equation}
\mathcal{L}_\text{F}=
N_\text{c} \tr\brk*{
-\partial_\mu \bar X \partial^\mu X 
-
\partial_\mu \bar Z \partial^\mu Z
+\xi^2 \bar X\bar Z X Z
}.
\label{eq:Lbi}
\end{equation} 
The new coupling $\xi:= g e^{-i\gamma_3/2}$ is kept fix in the limit.
Schematically, the route to the so-called \emph{bi-scalar fishnet theory} defined by $\mathcal{L}_\text{F}$ takes the form
\begin{equation}
\begin{tabular}{ccccc}
$\superN=4$ SYM&
\hspace{0mm}$\xrightarrow{\tiny X\,Y\to e^{i\gamma_j(\dots)}X 
 Y}$\hspace{-1mm}
&$\gamma$-Deformation&
$\xrightarrow[\xi= g e^{-i\gamma_3/2} \text{ fix}]{g\to 0,\,\, \gamma_{3}\to i\infty}$
&Fishnet Theory
\end{tabular}.
\label{eq:LimitRoute}
\end{equation}
Notably, the theory defined by \eqref{eq:Lbi} is non-unitary due to the chiral four-point interaction,
but it has the remarkable feature that correlation functions are in one-to-one correspondence with individual Feynman graphs of fishnet structure. 
In particular, these Feynman integrals (alias correlators) inherit a bosonic Yangian symmetry, i.e.\ the respective integrals are invariant under the level-one differential operators of the Yangian algebra $Y[\alg{so}(1,5)]$ \cite{Chicherin:2017cns,Chicherin:2017frs}:
\begin{equation}
\levo{J}^a
\includegraphicsbox[scale=.65]{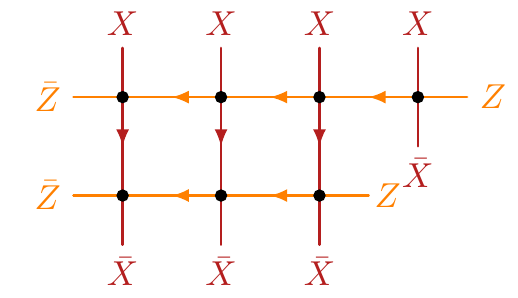}
=0.
\end{equation}
Next we discuss how the above integrability properties of massless Feynman integrals can be generalized to the massive situation.
\section{Massive Yangian Symmetry of Feynman Graphs}
\label{sec:MassYangFeyn}

In the massless case, $\mathcal{N}=4$ SYM theory was the starting point to deduce the integrability of massless Feynman graphs via the sequence  \eqref{eq:LimitRoute} .
Hence, it is suggestive to think about introducing masses into $\mathcal{N}=4$ SYM in order to follow a similar path. Masses are easily generated by giving a vacuum expectation value (VEV) to one of the scalar fields of $\mathcal{N}=4$ SYM theory:
$
\Phi  \to \langle \Phi\rangle +\Phi.
$
This leads to massive propagators with difference mass, i.e.\
\begin{equation}
\frac{1}{x_{jk}^{2}}\quad \to\quad \frac{1}{\hat x_{jk}^2}:= \frac{1}{{x_{jk}^2+(m_j-m_k)^2}}.
\label{eq:ToDiffMassProp}
\end{equation}
Here we use the notation $x_{jk}^\mu=x_j^\mu-x_k^\mu$ with $\mu=1,\dots, D$.
The mass can be interpreted as a fifth or $(D+1)$th dimension of the spacetime vector $x^{\hat \mu}$ with $\hat \mu=1,\dots,D+1$. For $D=4$ it plays the role of the radial direction in $\mathrm{AdS}_5$, i.e.\
$x^{\hat \mu=D+1}=m_j.$
There is a well known massive version of dual conformal symmetry in this massive phase, i.e.\ on the Coulomb branch of $\superN=4$ SYM theory, which is generated by the following differential operators \cite{Alday:2009zm}:
\begin{align}
\gen{P}^{\mu}_j &= -i \partial_{x_{j}}^{\mu}, 
\qquad\qquad
\gen{L}_j^{\mu \nu} = i x_j^{ \mu} \partial_{x_{j}}^{ \nu} - ix^{ \nu}_j \partial_{x_{j}}^{ \mu}, 
&
\gen{D}_j = -i \brk!{x_{j\mu} \partial_{x_j}^\mu + m_j \partial_{m_j} + \Delta_j},
\nonumber
\\
 \gen{K}^{ \mu}_j &= -2ix_j^{ \mu}\brk!{x_{j\nu}  \partial_{x_j}^\nu  +  m_j\partial_{m_j} + \Delta_j} +i (x^2_j + m^2_j)\partial_{x_{j}}^{ \mu}.
 \label{eq:MassDualLev0}
\end{align}
Importantly, however, no massive Yangian symmetry (integrability) is known on the Coulomb branch, i.e.\ the naive logic to deduce integrability properties of Feynman integrals via the sequence of steps \eqref{eq:LimitRoute} to the massive situation fails. We will come back to this approach in \Secref{sec:MassFishnet}.
%

An alternative route is to consider massive Feynman integrals directly, e.g.\ a diagram of the form
\begin{equation*}
\includegraphicsbox[scale=.8]{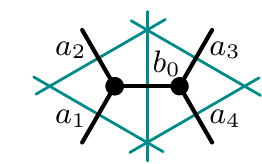}
\quad
=
\int \frac{\dd^D x_0 \dd^D x_{\bar 0}}
{
\hat x_{01}^{2a_1}
\hat x_{02}^{2a_2}
x_{0\bar 0}^{2b_0}
\hat x_{\bar 03}^{2a_3}
\hat x_{\bar 04}^{2a_4}
}.
\end{equation*}
Here the region momenta $x$ (black graph) can be related to ordinary momenta (green graph) via $p^{\mu}_j=x^{\mu}_j-x^{\mu}_{j+1}$.
Using their defining equation \eqref{eq:DefLev1}, it is straightforward to build Yangian level-one generators composed of densities of the above massive dual conformal (level-zero) generators $\levz^a$ as given in \eqref{eq:MassDualLev0}.
Here the level-one momentum generator for instance takes the form
\begin{equation}
\gen{\widehat P}^{\mu}
=\sfrac{i}{2} \sum_{j,k=1}^n 
\text{sign}(k-j)\brk!{\gen{P}_j^{\hat\mu} \gen{D}_k + \gen{P}_{j\nu} \gen{L}_k^{\nu \mu} }
 + \sum_{j=1}^n \Eval_j \gen{P}_j^{ \mu}.
\end{equation}
Acting with the so-constructed level-one generators on one- and two-loop Feynman integrals, one finds indeed invariance under  certain conditions.
Introducing generic propagator powers $a_j$, in particular the \emph{dual conformal condition (dcc)} $\sum_{k\in \text{vertex}} a_k=D$ plays an important role \cite{Loebbert:2020hxk,Loebbert:2020glj}:
\renewcommand{\arraystretch}{1.1}
\begin{center}
\begin{tabular}{|l | c |c| }
\hline
& 1 Loop & 2 Loops
\\\hline
& $I_n=\raisebox{1.4mm}{\includegraphicsbox[scale=.75]{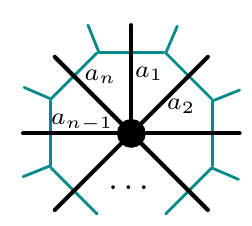}}$ & $I_{lr}=\includegraphicsbox[scale=.8]{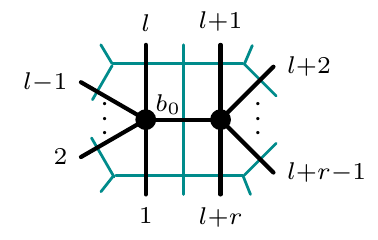}$
\\\hline
 {Level Zero}  &$\gen{J}^a I_n=0$ if dcc &  $\gen{J}^a I_{lr}=0$ if dcc
\\\hline
 {Level One} & $\levo{J}^a I_n=0$ always&  $\levo{J}^a I_{lr}=0$ if dcc, else $\levo{P}^\mu I_{lr}=0$ 
\\\hline
\end{tabular}
\end{center}
At higher loop orders it was conjectured in Ref.\ \cite{Loebbert:2020hxk} that 
all planar Feynman graphs, which are cut along a closed contour from one of the three regular tilings of the plane, have massive Yangian symmetry if all internal propagators are massless, while external 
propagators can be massive or massless.
This conjecture is motivated by the fact that it agrees with the findings on regular tilings in the massless limit \cite{Chicherin:2017cns,Chicherin:2017frs} and it is supported by numerical evidence for certain example diagrams.


An interesting picture arises when one translates the Yangian level-one momentum generator from region ($x$) to ordinary momentum ($p$) space related via $p^{\mu}_j=x^{\mu}_j-x^{\mu}_{j+1}$. Here the level-one momentum $\levo{P}^\mu$ is mapped to a differential operator $\gen{\bar K}^\mu$ which forms part of a \emph{massive extension} of the conformal algebra in momentum space \cite{Loebbert:2020hxk,Loebbert:2020glj}:
\begin{align}
\gen{\bar P}_{j}^\mu &= p_j^\mu \, , 
\qquad
\gen{\bar L}_{j}^{\mu\nu} =  p_j^\mu \partial_{p_j}^{\nu} -  p_j^\nu \partial_{p_j}^{\mu},
\qquad
\gen{\bar D}_{j} = p_{j\nu} \partial^{\nu}_{p_j}+ \sfrac{m_j \partial_{m_j}+m_{j+1} \partial_{m_{j+1}}}{2} + \bar{\Delta}_j,
\nonumber
\\
\gen{\bar K}_{j}^\mu 
&= p_j^\mu \partial_{p_j}^2
-2\brk[s]2{p_{j\nu} \partial_{p_j}^\nu+\sfrac{m_j \partial_{m_j}+m_{j+1} \partial_{m_{j+1}}}{2} + \bar{\Delta}_j }\partial_{p_j}^{\mu}.
\label{eqn:MomConfMassRep}
\end{align}
This new massive generalization extends the previous massive \emph{dual} conformal symmetry~\cite{Alday:2009zm}.

\section{Yangian Bootstrap for All-Mass $n$-Gons}
\label{sec:YangianBootstrap}

As an interlude, we present a useful bootstrap application of the above Yangian symmetry, cf.\ Ref.\ \cite{Loebbert:2019vcj,Loebbert:2020glj,Corcoran:2020epz}. We ask the question how constraining the symmetry is for generic one-loop integrals with conformal condition  $D=\sum_{j}a_j$ and all propagators being massive.
Level-zero symmetry implies that, up to an overall weight, the integral is a function of a massive generalisation of the conformal cross ratios:
$
 u_{ij} =(x_{ij}^2+(m_i-m_j)^2)/(-4m_i m_j).
$
The new level-one symmetry then implies sets of partial differential equations for this function
$\phi(u_{ij})$ whose cardinality depends on the number of variables. The study of examples with small numbers of legs leads to a natural conjecture for $n$-point diagrams:
\begin{itemize}
\item[] $n=2$: fixed by symmetries (Gau{\ss}' hypergeometric function)
\begin{align*}
I_2 = 
\includegraphicsbox[scale=.9]{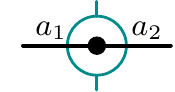}
=\frac{\pi^{D/2}\Gamma_{D/2}}{\Gamma_D m_1^{a_1}m_2^{a_2}} 
\GG{2}{1}{a_1,a_2}{(D+1)/2}{u}.
\end{align*}

\item[] $n=3$: fixed by symmetries (Srivastava's triple hypergeometric function)
\begin{equation}
I_3=
\includegraphicsbox[scale=.85]{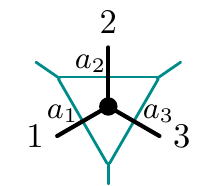}=
\frac{\pi^{D/2}\Gamma_{D/2}}{\Gamma_D m_1^{a_1}m_2^{a_2}m_3^{a_3}}
H_{C}(u,v,w).
\end{equation}

\item[] $n=$generic: `experiments' yield natural conjecture   \cite{Loebbert:2020glj}
\begin{equation}
I_n=\raisebox{1.4mm}{\includegraphicsbox[scale=.85]{FigNStarPropWeightsDots.pdf}}
= \frac{\pi^{D/2}\Gamma_{D/2}}{\Gamma_D\prod_{j=1}^n{m_j^{a_j}}}\sum_{k_{ij}=0}^\infty \frac{\prod_{j=1}^n (a_j)_{\sum_{\alpha\in B_{n|j}}  k_{\alpha}}}{(\sfrac{D+1}{2})_{\sum_{\alpha\in B_n}k_{\alpha}}} \prod_{\alpha\in B_n} \frac{u_\alpha^{k_\alpha}}{k_\alpha!}.
\end{equation}
\end{itemize}
This  $n$-point conjecture was recently confirmed in Ref.\ \cite{Ananthanarayan:2020xpd}.

\section{From Feynman Graphs to Massive Fishnet Theory }
\label{sec:MassFishnet}

Having motivated the Yangian symmetry of massless Feynman integrals via the relation to a massless fishnet theory, an obvious question is whether the Yangian symmetry of masssive Feynman integrals has a similar origin.
In other words: Can we define a massive version of the fishnet theory?
The simplest idea would be to take the massless fishnet theory and to introduce mass via spontaneous symmetry breaking. This line of thought was explored in Ref.\ \cite{Karananas:2019fox,Loebbert:2020tje} with the result that masses enter into the propagators of the theory as products, i.e.\ in the form $p^2+m_jm_k$. Hence, the resulting Feynman graphs do not correspond to the Yangian-invariant integrals with difference-mass propagators \eqref{eq:ToDiffMassProp}.
An alternative idea is to first introduce mass in $\superN=4$ SYM theory and to then take a double-scaling limit on the resulting Coulomb branch \cite{Loebbert:2020tje}. In fact, this approach yields the desired difference-mass propagators $p^2+(m_j-m_k)^2$. However, taking a double-scaling limit on the Coulomb branch of $\superN=4$ SYM theory requires some care.
In particular, a critical step is the $\gamma$-deformation $\mathcal{P}_\gamma$, which a priori is only well-defined on R-symmetry singlets.
A solution, motivated by the resulting limit theory, is to average over all ways to break up traces in Lagrangian:
\begin{equation}
Q: \tr(\Phi_1\Phi_2\dots \Phi_n) \mapsto
\sfrac{1}{n} \big (\Phi_1\Phi_2\dots \Phi_n
+  \Phi_2\Phi_3\dots \Phi_1
+\dots
\big).
\end{equation}
Then we define the $\gamma$-deformed Coulomb-branch Lagrangian as
$
\mathcal{L}_\text{Coul}^{\gamma}
=
Q^{-1}\mathcal{P}_\gamma Q\mathcal{L}_\text{Coul}.
$
Importantly, application of $Q^{-1}\mathcal{P}_\gamma Q$ is now a well defined mathematical operation even if the R-symmetry of $\mathcal{L}_\text{Coul}$ is broken.
Following the massless procedure, the most restrictive double-scaling limit then results in a \emph{massive fishnet theory} defined by the following Lagrangian \cite{Loebbert:2020tje}:
\begin{align}
\mathcal{L}_\text{MF}=
&N_\text{c} \tr\brk*{
-\partial_\mu \bar X \partial^\mu X 
-
\partial_\mu \bar Z \partial^\mu Z
+\xi^2 \bar X\bar Z X Z
}
\nonumber\\
&-  N_\text{c}(m_a - m_b)^2 {X^a}_b{\bar X^b}_a
- N_\text{c}(m_a - m_b)^2 {Z^a}_b{\bar Z^b}_a.
\label{eq:LMassFishnet}
\end{align} 
Planar amplitudes in the respective theory are in one-to-one correspondence with Yangian-invariant massive Feynman integrals. As sketched in \Secref{sec:MassYangFeyn}, this invariance is proven at one and two loops and conjectural at higher loop orders. In this sense the above massive fishnet theory is integrable.
%

The relations between the different theories are summarized in the following figure:%
\begin{center}
\includegraphicsbox{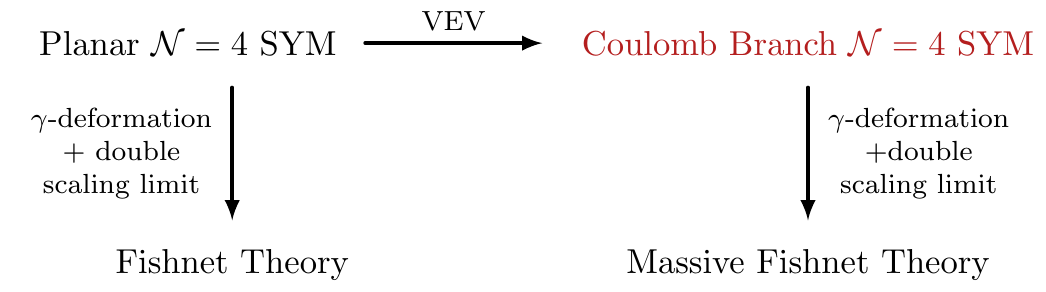}
\smallskip
\end{center}
This diagram with three integrable corners (black) leads to a natural question:
Is planar $\superN$=4 SYM theory on the Coulomb branch integrable?


\acknowledgments

We are grateful to Dennis M\"uller and Hagen M\"unkler for collaborations on the main topics discussed in this contribution.  The work of FL is funded by the Deutsche Forschungsgemeinschaft (DFG, German Research Foundation)--Projektnummer 363895012. JM is  supported  by  the  International  Max  Planck  Research  School  for  Mathematical  and Physical Aspects of Gravitation, Cosmology and Quantum Field Theory.
\bibliographystyle{JHEP}
\bibliography{ProceedingsEPSHEP2021}

\end{document}